\documentclass{jpsj3}
\usepackage{txfonts}
\usepackage{color}
\usepackage{braket}
\usepackage{bm}

\title{Universal Behavior of Magnetoresistance in Organic Dirac Electron Systems}

\author{Ryotaro Kobara$^1$, Shin Igarashi$^1$, Yoshitaka Kawasugi$^1$, Ryusei Doi$^2$, Toshio Naito$^2$, Masafumi Tamura$^3$, Reizo Kato$^4$, Yutaka Nishio$^1$, Koji Kajita$^1$, and Naoya Tajima$^1$\thanks{naoya.tajima@sci.toho-u.ac.jp}}

\inst{
$^1$Department of Physics, Toho University, Funabashi, Chiba 274-8510, Japan\\ 
$^2$Graduate School of Science and Engineering, Ehime University, Matsuyama 790-8577, Japan \\
$^3$Department of Physics, Faculty of Science and Technology, Tokyo University of Science, Noda, Chiba 278-8510, Japan\\
$^4$RIKEN, Hirosawa 2-1, Wako-shi, Saitama 351-0198, Japan
}

\abst{
 In-plane magnetoresistance for organic massless Dirac electron system (OMDES) $\alpha$-(BEDT-TTF)$_2$I$_3$ and $\theta$-(BEDT-TTF)$_2$I$_3$ in addition to possible candidates of the OMDES $\alpha$-(BETS)$_2$I$_3$ and $\alpha$-(BEDT-STF)$_2$I$_3$, was investigated under hydrostatic pressure. We have found the universal behavior of the in-plane magnetoresistance under a low magnetic field perpendicular to two-dimensional plane. As for $\alpha$-(BEDT-TTF)$_2$I$_3$, the universality was examined with the parameters of temperature, magnetic field and its direction. We suggest that the universal magnetoresistance behavior is found even for the gapped state of $\alpha$-(BEDT-TTF)$_2$I$_3$ under intermediate pressure, when the thermal energy exceeds the gap.
}

\allowdisplaybreaks

\newcommand{\be}{ \begin{equation}}
\newcommand{\ee}{ \end{equation}}

\usepackage{color}

\usepackage{ulem}

\begin{document}
\maketitle

A variety of materials with Dirac electrons have emerged, thanks to recent progress in the physics of Dirac electrons in a solid following the discovery of graphene\cite{geim2005, kim2005}. However, examples of massless Dirac electron systems (MDES), where the Fermi level is at the Dirac point (DP), are still limited. Two-dimensional (2D) OMDES, $\alpha$-(BEDT-TTF)$_2$I$_3$ under high pressure and related materials, have been explored to uncover physical phenomena at DP\cite{kajita2014}. In contrast to graphene, these materials have an experimental advantage that their Dirac electron behavior can be observed in bulk crystals.

In a magnetic field perpendicular to a 2D plane, the relativistic Landau level (LL) structure in MDES is expressed as $E_{N}={\rm sgn}(N)\sqrt{2e\hbar v_{F}^2|N||B|}$, where $N$ is the Landau index and $B$ is the magnetic field strength. The Fermi velocity, $v_F$, in $\alpha$-(BEDT-TTF)$_2$I$_3$ under pressure, for example, is approximately $3.5\times10^4$ m/s\cite{kajita2014}. Of importance is the appearance of the $N=0$ level (LL$_0$) at DP when magnetic fields are applied normal to the 2D plane. Because this level is at the Fermi energy irrespective of the field strength, most of the mobile carriers are in LL$_0$ for $k_{\rm B}T \ll E_{\pm 1}$. Such a situation is called the quantum limit. The OMDES $\alpha$-(BEDT-TTF)$_2$I$_3$ under high pressure has thus opened doors to the physics of LL$_0$\cite{osada2008, tajima2009, konoike2013}. 

In this letter, we unveil a new common magnetoresistance behavior at a low magnetic field normal to the 2D plane and a high temperature ($k_{\rm B}T \gg E_{\pm 1}$), as the quantum limit is approached in OMDES. The magnetoresistance reflects how the electrons are distributed over $N \neq 0$ components under a magnetic field, as the half-occupied LL$_0$ exhibits constant conductivity $e^2 / \pi h$ so as not to contribute magnetoresistance. The characteristics of OMDES, as found in this work, have never been observed in conventional metals and semiconductors. 

In-plane magnetoresistance for four kinds of OMDES, $\alpha$-(BEDT-TTF)$_2$I$_3$ ($\alpha$-ET), $\alpha$-(BETS)$_2$I$_3$ ($\alpha$-BETS), $\theta$-(BEDT-TTF)$_2$I$_3$ ($\theta$-ET), and $\alpha$-(BEDT-STF)$_2$I$_3$ ($\alpha$-STF), was investigated under hydrostatic pressure. 

A single crystal of $\alpha$-ET consists of alternately stacked conductive layers of BEDT-TTF molecules and insulating layers of I$_3$$^-$ anions\cite{bender1984}. Under ambient pressure, this material exhibits metallic properties down to 135 K, where it undergoes phase transition to a charge-ordered insulator with a horizontal charge stripe pattern\cite{kino1995,takano2001}. A high hydrostatic pressure above 1.5 GPa suppresses the metal-insulator transition, giving rise to the MDES\cite{kajita2014}. 

$\alpha$-BETS\cite{inokuchi1995} and $\alpha$-STF\cite{naito1997, naito2020, naito2020b} are isostructural to $\alpha$-ET. The sulfur (S) atoms in BEDT-TTF molecules are partially replaced by selenium (Se) atoms in BETS and BEDT-STF molecules, as shown in the insets of Figs. 1(b) and (d). Under ambient pressure, $\alpha$-BETS\cite{inokuchi1995} and $\alpha$-STF\cite{naito1997, naito2020, naito2020b} undergo phase transition to an insulator at approximately 50 K and 80 K,
respectively. 
The origins of these insulating phases are not understood yet. On the other hand, these materials under hydrostatic pressure exhibit peculiar magnetotransport phenomena similar to those of $\alpha$-ET\cite{inokuchi1993, tajima2006}.
 Therefore, they presumably belong to the OMDES.

$\theta$-ET, on the other hand, is a typical 2D metal with large Fermi surfaces at ambient pressure. This material undergoes phase transition to the MDES at 0.5 GPa and ambient temperature\cite{tamura1997, tajima2004, miyagawa2010, kajita1992}. 

In this letter, we propose the universality of the in-plane magnetoresistance behavior in the four salts. In all four cases, carrier mobility is enhanced at high pressure. In order to understand the basic properties of Dirac electrons, the materials should be set as far as possible from the insulating phase, so that all the samples are pressurized up to 1.7 GPa. In particular, $\alpha$-ET is examined in detail with the magnetic field ($B$) and its direction, the temperature ($T$), and the pressure ($p$) as parameters.

We show the $T$ dependence of the in-plane resistivity ($\rho$) of $\alpha$-ET, $\alpha$-BETS, $\theta$-ET, and $\alpha$-STF under $B$ in Figs. 1(a), (b), (c), and (d), respectively. There are reports on the $\rho$ of $\alpha$-ET under $B$\cite{kajita2014, inokuchi1993, kajita1992}. The field-induced large stepwise change in $\rho$ shown in Fig. 1(a) was reproduced by the linear response theory in terms of LL$_0$ and its spin splitting\cite{mirinari2010, proskurin2015}. Under 5 T, for example, the shoulder of $\rho$ at around 10 K indicates that the system is in the quantum limit dominated by LL$_0$ carriers. The spin-split LL$_0$ gives rise to the increase in $\rho$ at low $T$. The behavior of $\rho$ for $\alpha$-BETS and $\theta$-ET is almost identical with that for $\alpha$-ET as shown in Figs. 1(b) and (c).

At low $B$ and/or high $T$, where $k_{\rm B}T \gg |E_{\pm 1}|$, the transverse magnetoresistance ($\Delta \rho/\rho_0$) of the three materials is reproduced by $\Delta \rho/\rho_0 \propto B^{3/2}$, where $\Delta \rho=\rho-\rho_0$ and $\rho_0$ is the resistivity at $B=0$.  Surprisingly, $[\Delta \rho/\rho_0]B^{-3/2}$ in the three systems lies on the same line, as shown in Fig. 1(e),
indicating that the three materials belong to the same class.
The experimental formula is written as 
\begin{equation}
\label{eq1}
\Delta \rho/\rho_0 = a B^{3/2}T^{\beta}, 
\end{equation}
where $a\simeq 1.8 \times 10^6$ T$^{-3/2}$$\cdot$K$^5$ and $\beta \simeq -5$. To our knowledge, the $B^{3/2}$ law of $\Delta \rho/\rho_0 $ is the first finding and no theoretical explanation or prediction for this has been given yet. Nevertheless, we can relate the temperature dependence of magnetoresistance to the carrier lifetime, $\tau$. Generally, $\Delta \rho/\rho_0 \propto \tau^2$ for a constant field, and temperature dependence of $\tau$ can be estimated, for example, from the carrier mobility, $\mu$. Thus, $\tau \propto T^{\beta/2}$ should be compared with the effective carrier mobility expressed experimentally as $\mu \propto T^{\gamma}$ where $\gamma =\beta/2 \simeq -2.3$,\cite{kajita2014, tajima2006, tajima2004} which is close to $\beta/2 \simeq -2.5$ found in this work. 
%The origin of temperature dependence of $\tau \propto T^{-5/2}$ has not been opened further discussions. 

The good agreement of $\Delta \rho/\rho_0 \propto B^{3/2}$ in the three materials indicates the universality of $\Delta \rho/\rho_0$ behavior. In general metals or semiconductors, $\Delta \rho/\rho_0 \propto B^2$ at low $B$. Linear magnetoresistance, on the other hand, is recognized in a topological insulator and inorganic MDES\cite{tang2011, feng2015} with the Fermi energies deviating from the DPs. We surmise that the observed $B^{3/2}$ law of $\Delta \rho/\rho_0$ involves a new physics of LL$_0$. 
We tentatively think of a possible explanation for this behavior. Naively speaking, the $N \neq 0$ components, giving the usual $B^2$ contribution, efficiently deviate from the range, $E_0 \pm k_{\rm B} T$, as $B^{-1/2}$ at a low field as a result of $E_N \propto B^{1/2}$. Thus, we may observe the $B^{-1/2} B^2 = B^{3/2}$ behavior of magnetoresistance. 

Next, we will discuss the effects of twinning in $\theta$-ET or disorder in $\alpha$-STF on the $\Delta \rho/\rho_0 \propto B^{3/2}$ law. 
$\theta$-ET is known to show a  twinning nature due mainly to the two possible ordered arrangements of the I$_3$$^{-}$ anions\cite{kobayasji1986, tamura1990, terashima1994}. 
In fact, cleavage along the 2D plane occurs sometimes due to the release of $p$ after measurements. However, twining does not appear to affect $\Delta \rho/\rho_0$. Clear $B^{3/2}$ behavior is observed in this system, as shown in Fig. 1(c). This is well understood when one consider that $\theta$-ET is an ideally 2D system, where the insulating sheets of anions hardly affect the electrical properties.
On the other hand, the situation is different in a disordered system $\alpha$-STF, as shown in Figs. 1(d) and (e). We expect that the orientational disorder due to asymmetric structure of BEDT-STF molecules\cite{naito2020, naito2020b, inokuchi1993} would give rise to a high damping rate in this system. The high damping rate broadens the width $\Gamma$ of the Landau levels ($\Gamma = h/\tau$) and then obstructs the formation of the quantum limit. This leads to a small $\Delta \rho/\rho_0$ and a less pronounced shoulder of $\rho$, as shown in Fig. 1(d). Nevertheless, it is surprising to obtain the universal relation $\Delta \rho/\rho_0 \propto B^{3/2}$ in Fig. 1(e). The $T$ dependence of $[\Delta \rho/\rho_0]B^{-3/2}$ under different magnetic fields is reproduced by a single curve. 

The disorder in the 2D plane, on the other hand, will affect the $T$ dependence of the carrier mobility to alter the values of $\alpha$ and $\beta$ in Eq. (1). The experimental formula of $\Delta \rho(B)/\rho(0)$ at low $B$ and high $T$ for $\alpha$-STF is expressed as 
\begin{equation}
\label{eq2}
\Delta \rho/\rho_0 = b B^{3/2}T^{-5/2}, 
\end{equation}
where $b\simeq 95$ T$^{-3/2}$K$^{5/2}$, as shown in Fig. 1(e).

In $\alpha$-ET, the energy gap (Dirac mass gap $E_D$) can be modified by controlling the pressure within $p=1.05-1.7$  GPa and $T<65$ K. This enables us to control the Dirac mass in this system. The $T$ dependence of $\rho$ under selected $p$ is shown in Fig. 2(a). At $p=$1.05 and 1.15 GPa, this material undergoes phase transition to a correlated insulator at approximately 55 K and 35 K, respectively, where the system is expected to be massive (gapped)\cite{beyer2016, ohki2019}. This material at $p\geq 1.3$ GPa, on the other hand, has the ground state of the MDES. 

The magnetoresistance behavior $\Delta \rho/\rho_0 \propto B^{\alpha}$ is observed even at $p=1.05-1.7$ GPa. We plot the exponent part $\alpha$ as a function of $T$ in Fig. 2(b). The value, $\alpha \simeq 3/2$, is independent of $p$ and $T$ for $T\geq17$ K. For example, we show the $B$ dependence of $\Delta \rho/\rho_0$ at $p=$1.05 GPa in Fig. 3. We find that the relation $\Delta \rho/\rho_0 \propto B^{3/2}$ holds for $T\geq17$ K even in the massive case. 
The energy gap $E_D/k_B$ at $p=$1.05 and 1.15 GPa is estimated to be approximately 12 K and 8 K, respectively, from the activation energy of $\rho$ for $T<10$ K. $E_D < k_BT\sim \Gamma$ under $B$ can be assumed when $\Delta \rho/\rho_0$ follows $\Delta \rho/\rho_0 \propto B^{3/2}$ law. If the lowest LL above the gap and the highest LL below the gap are almost equally occupied, their contribution to magnetoresistance disappears so as to retain the $B^{3/2}$ behavior similarly to the massless case.

At $p=$1.7 GPa, we examine the effects of Landau quantization for the magnetic field applied in two directions, $B \parallel$2D-plane and $B\perp$2D-plane (Fig. 4).  As shown in Figs. 4(a) and (b), the transverse magnetoresistance for $B\parallel$2D-plane obeys $\Delta \rho/\rho_0 \propto B^2$ for $T \leq$ 4 K as in an ordinary metal, rather than $\Delta \rho/\rho_0 \propto B^{3/2}$, because LL$_0$ does not exist in this case.

Figures 4(c) and (d) show $\Delta \rho/\rho_0$ for $B\perp$2D-plane. We find again a shoulder of $\Delta \rho/\rho_0$, which is a sign of the quantum limit in Fig. 4(c). At 4 K, for example, the critical field $B_s$ is approximately 0.2 T\cite{tajima2009}.  It increases with increasing temperature as $B_s \propto {T^2}$\cite{sugawara2010}. What is significant is that $\Delta \rho/\rho_0 \propto B^{\alpha}$ with $\alpha \simeq 3/2$ for $|B|<B_s$ and $T>2.2$ K, as shown in Fig. 4(d). At 30 K, we find the universal relation at $|B|<5$ T. Those results indicate that the behavior $\Delta \rho/\rho_0 \propto B^{3/2}$ is associated with LL$_0$. 

We would like to make a brief comment on the low temperature behavior.
The linear magnetoresistance behavior is observed at $|B|\leq 10$ mT, $p= 1.05-1.7$ GPa, and $T < 2$ K in Figs. 3 and 4. Linear magnetoresistance independent of the field direction shown in Fig. 4 suggests that the origin is different from that in the topological insulators and the inorganic MDES. Recently, in the same $T$ and $B$ ranges, Konoike found a diamagnetic phase in this system\cite{konoike2019}. We suppose that the linear magnetoresistance may be related to the diamagnetic phase. On the other hand, the negative magnetoresistance at $|B|>0.5$ T ($B \parallel $2D-plane) and 0.5 K may be associated with the ferrimagnetic state discovered by Hirata et al \cite{hirata2016}. Both issues will be reported separately.

In conclusion, we have found experimentally that the in-plane magnetoresistance in the four salts (including the twin and disordered systems) obeys the $B^{3/2}$ law under a low magnetic field and at a high temperature near the quantum limit, which suggests that both $\alpha$-BETS and $\alpha$-STF also belong to the OMDES as anticipated. 
For the three materials with little disorder in the 2D plane, the $[\Delta \rho/\rho_0]B^{-3/2}$ curves as a function of $T$ were quantitatively identical to each other. This indicates that the $B^{3/2}$ law is universal and specific to the massless Dirac electrons with the Fermi energy at the DPs approaching the quantum limit. This universality of the $B^{3/2}$ behavior is extended by the observation of this behavior even in the weakly massive Dirac electron states of  $\alpha$-ET at $p=1.05$ and 1.15 GPa when the thermal energy exceeds the energy gap. 

\acknowledgments{This work was supported by MEXT/JSPJ KAKENHI under Grant No 16H06346.}

%-----------------------------------------------

%------- Figure

\newpage
\large{two column}

%=========================Fig. 1==========================
\begin{figure}
\begin{centering}
\includegraphics[trim=10 10 20 0, width=11cm, angle=270,clip]{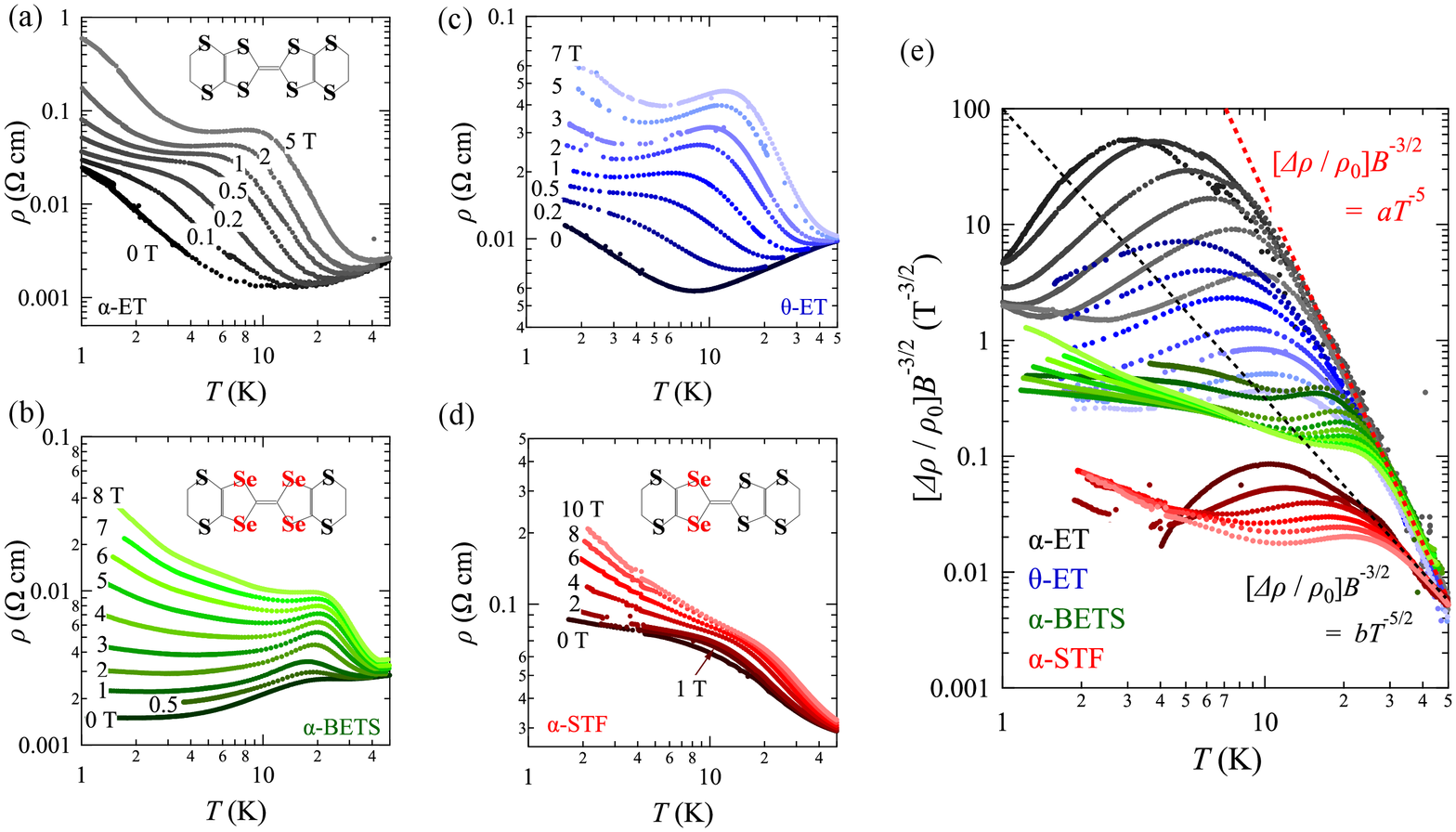}
\caption{\label{fig1}(Color online) 
Temperature dependence of $\rho$ under various $B$ for (a) $\alpha$-ET, (b) $\alpha$-BETS, (c) $\theta$-ET, and (d) $\alpha$-STF. All $\rho$ data were measured at 1.7 GPa. Insets in (a), (b), and (d) shows the BEDT-TTF, BETS, and BEDT-STF molecules. (e) Temperature dependence of $[\Delta \rho/\rho_0]B^{-3/2}$ of the four materials. Broken lines indicate the universal behavior as the power law expressed in Eqs. (1) and (2).
}
\end{centering}
\end{figure}
%=======================================================

\newpage
\large{one column}

%=========================Fig. 2==========================
\begin{figure}
\begin{centering}
\includegraphics[width=7cm]{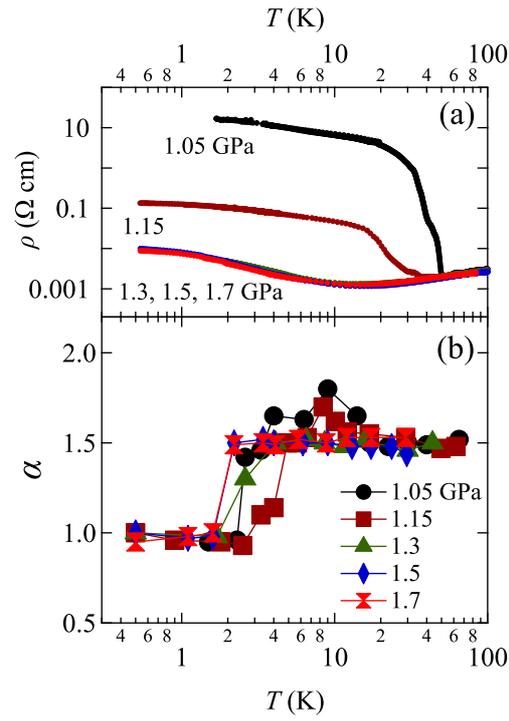}
\caption{\label{fig2}(Color online)
Temperature dependence of (a) $\rho$ and (b) the exponent part $\alpha$ of $\Delta \rho/\rho_0 \propto B^{\alpha}$ for $\alpha$-ET under several pressure above 1.05 GPa.
}
\end{centering}
\end{figure}
%=======================================================

\newpage
\large{one column}

%=========================Fig. 3==========================
\begin{figure}
\begin{centering}
\includegraphics[trim=10 20 50 0, width=6cm, angle=270,clip]{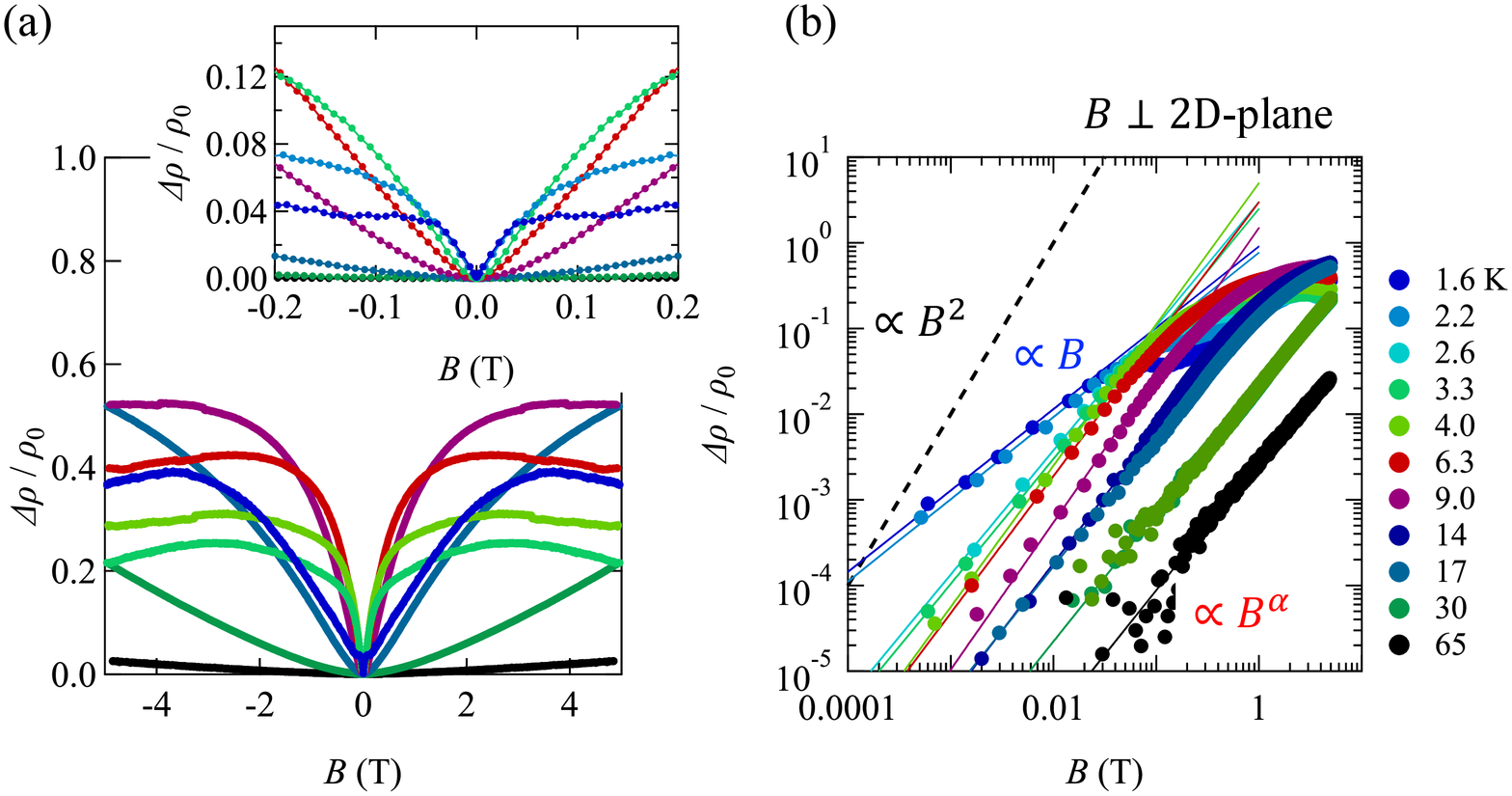}
\caption{\label{fig3}(Color online)
Magnetic field dependence of $\Delta \rho/\rho_0$ for $\alpha$-ET at $p=1.05$ GPa and $T<65$ K. The magnetoresistance for the magnetic field perpendicular to 2D plane is shown by linear (a) and log-log  (b) scales. In (b), each set of data at low $B$ is fit by the function $\Delta \rho/\rho_0 \propto B^{\alpha}$, where the exponent part $\alpha$ is plotted in Fig. 2(b). The broken line shows a curve of $\Delta \rho/\rho_0 \propto B^{2}$. The inset in (a) shows $\Delta \rho/\rho_0$ for $|B|<0.2$ T.
}
\end{centering}
\end{figure}
%=======================================================

\newpage
\large{one column}

%=========================Fig. 4==========================
\begin{figure}
\begin{centering}
\includegraphics[trim=50 20 10 0, width=8.5cm]{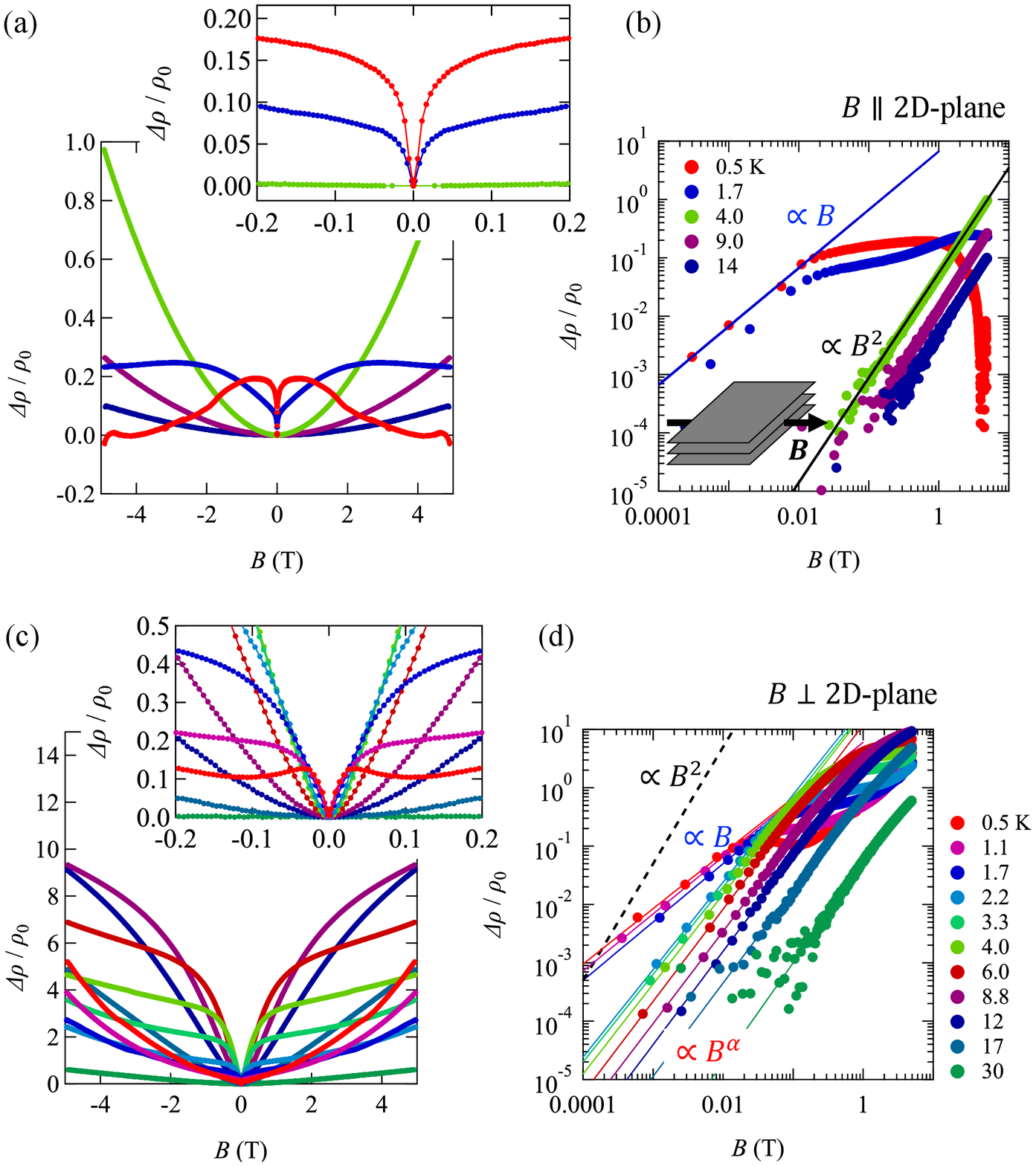}
\caption{\label{fig4}(Color online)
Magnetic field dependence of $\Delta \rho/\rho_0$ for $\alpha$-ET at $p=1.7$ GPa and $T<30$ K. The transverse magnetoresistance for $B$ parallel to 2D plane is shown by the linear (a) and the log-log (b) scales. The magnetoresistance at 4.0 K, 9.0 K, and 14 K, expressed as $\Delta \rho/\rho_0 \propto B^{2}$. Figures of the linear (c) and the log-log (d) scales are the magnetoresistance for $B$ perpendicular to 2D plane. In (d), each set of data at low $B$ is fit by the function $\Delta \rho/\rho_0 \propto B^{\alpha}$, where the exponent part $\alpha$ is plotted in Fig. 2(b). The broken line shows a curve of $\Delta \rho/\rho_0 \propto B^{2}$. The insets in (a) and (c) show $\Delta \rho/\rho_0$ for $|B|<0.2$ T. The linear magnetoresistances observed under $|B|<10$ mT at $T<2$ K are field-angle-independent.
}
\end{centering}
\end{figure}
%=======================================================

\end{document}